\documentclass[aps,pra,twocolumn,nofootinbib]{revtex4-2}
\usepackage{amsmath,amssymb,amsthm}
\usepackage{graphicx}
\usepackage[hidelinks]{hyperref}
\usepackage{parskip}
\usepackage[inline]{enumitem}
\usepackage{tikz}
\usetikzlibrary{positioning,arrows.meta,calc}
\usepackage[capitalise]{cleveref}
\newcommand{\Paragraph}[1]{{\smallskip\noindent \bf #1}}
\newcommand{\SubParagraph}[1]{{\smallskip\noindent \em #1}}

\newcommand{\GL}[1]{\mathsf{GL}(#1,2)}
\newcommand{\CNOTDistance}{\mathsf{D_{CNOT}}}
\newcommand{\XORDistance}{\mathsf{D_{XOR}}}
\newcommand{\Nats}{\mathbb{N}}
\newcommand{\Eval}{\mathsf{eval}}
\newcommand{\Inputs}{\mathcal{I}}

\newcommand{\Gates}{\mathcal{G}}
\newcommand{\Field}{\mathbb{F}_2}
\newcommand{\NP}{\textsf{NP}}
\newcommand{\EdgeForms}{\mathcal{F}}
\newcommand{\RevEdgeForms}{\mathcal{R}}

\theoremstyle{plain}            
\newtheorem{theorem}{Theorem}
\newtheorem{proposition}[theorem]{Proposition}  
\newtheorem{lemma}[theorem]{Lemma}

\theoremstyle{definition}       

\theoremstyle{remark}

\AddToHook{env/proposition/begin}{\crefalias{theorem}{proposition}}
\AddToHook{env/lemma/begin}{\crefalias{theorem}{lemma}}
\AddToHook{env/corollary/begin}{\crefalias{theorem}{corollary}}
\AddToHook{env/definition/begin}{\crefalias{theorem}{definition}}

\newlist{compactenum}{enumerate}{3} 
\setlist[compactenum]{nosep, leftmargin=*}
\setlist[compactenum,1]{label=(\arabic*), ref=(\arabic*)}
\setlist[compactenum,2]{label=(\alph*), ref=(\arabic{compactenumi}\alph*)}
\setlist[compactenum,3]{label=(\roman*), ref=(\arabic{compactenumi}\alph{compactenumii}\roman*)}
\crefname{compactenumi}{Step}{Steps}
\crefname{compactenumii}{Step}{Steps}
\crefname{compactenumiii}{Step}{Steps}

\newlist{compactitem}{itemize}{3} 
\setlist[compactitem]{nosep, leftmargin=*}
\setlist[compactitem,1]{label=\textbullet}
\setlist[compactitem,2]{label=\textendash}
\setlist[compactitem,3]{label=\textasteriskcentered}

\newlist{invariants}{enumerate}{1}
\setlist[invariants]{label=(\arabic*), ref=(\arabic*), nosep, leftmargin=*}
\crefname{invariantsi}{Item}{Items}

\newcounter{subfig}[figure]

\crefname{subfig}{Fig.}{Figs.}

\begin{document}
\title{On the Hardness of Optimal CNOT Circuits}
\author{Andreas Pavlogiannis}
\affiliation{Aarhus University, Denmark\\\href{mailto:pavlogiannis@cs.au.dk}{pavlogiannis@cs.au.dk}}

\begin{abstract}
This note revisits the computational problem of optimal CNOT synthesis:~given an invertible binary matrix $A\in \GL{d}$ over the field of two elements and a budget $K$,
decide if $A$ can be implemented with $\leq K$ CNOT gates.
The $\NP$-completeness of this problem was recently published independently by Acuaviva, Acuaviva, and Acuaviva (arXiv:2608.03825) and by Li and Guan (arXiv:2609.04160).
This note gives a new proof, with three features.
First, $\NP$-completeness holds even over linear oracles, i.e., matrices of the form $(\mathbf{x},\mathbf{y})\mapsto(\mathbf{x},\mathbf{y}+F\mathbf{x})$,
which are the standard reversible embedding of a linear function $F$.
These are involutions, and seemingly simpler than the families of the two previous works, which both have order four.
Second, hardness persists when SWAP gates are free and qubits may be permuted (and this holds under any connected topology).
Third, the exposition is arguably simpler to follow.
Along the way, it gives a proof of the XOR-circuit lower bound of Boyar, Matthews, and Peralta [J.~Cryptology 26 (2013)] that is, perhaps, simpler than existing ones.
\end{abstract}

\maketitle

I publish this paper with some diffidence, since
(i)~its main result is known from prior works~\cite{Acuaviva2026,Li2026}, and
(ii)~as in those works, AI was involved in the process.
Specifically, I instructed Claude Opus 5.5 to obtain $\NP$-hardness for a specific family of linear oracles, and to simplify a proof of~\cite{Boyar2013} using localized arguments based on gate elimination.
I then simplified and polished the resulting proofs, and added expository text explaining the underlying intuition.
Naturally, I take responsibility for the paper as a whole,
which may be of interest for a few reasons:
\begin{enumerate*}[label=(\roman*)]
\item hardness holds even for linear oracles,
\item also even with free swaps and qubit permutations, and thus, in this case, under any connected communication topology, and
\item a short, intuitive exposition, including a simpler proof of the lower bound of~\cite{Boyar2013}.
\end{enumerate*}

\section{Preliminaries}\label{sec:preliminaries}

We introduce basic definitions on XOR/CNOT circuits, and a promise version of the vertex cover problem that remains $\NP$-hard, serving as the source of our reduction.

\Paragraph{XOR circuits.}
An \emph{XOR circuit} is a pair $(\Inputs, \Gates)$, where
\begin{enumerate*}[label=(\roman*)]
\item $\Inputs$ is a finite set of inputs, and
\item $\Gates=(g_\ell \gets a_\ell + b_\ell)_{\ell\in[s]}$ is a sequence of $s$ gates,
  where each $g_\ell$ is a fresh variable and
  $a_\ell, b_\ell \in \Inputs \cup \{g_{\ell'}\colon \ell' < \ell\}$.
\end{enumerate*}
The \emph{variables} of the circuit are its inputs and its gates.
The \emph{expression} computed by a variable is defined recursively by
$\Eval(x)=x$ for $x\in\Inputs$ and $\Eval(g_\ell)=\Eval(a_\ell)+\Eval(b_\ell)$
(where addition is over $\Field$, so two occurrences of an input cancel).
Fixing an order $x_1,\dots,x_d$ on $\Inputs$, we identify each expression with the row vector in $\Field^{1\times d}$
whose $j$-th entry is $1$ iff $x_j$ occurs in it.
The circuit \emph{implements} a matrix $A$ with $d$ columns if every row of $A$ equals $\Eval(v)$ for some variable $v$;
that is, the circuit computes every row of $A$, and possibly other expressions.
Matrices act on input column vectors as $\mathbf{v}\mapsto A\mathbf{v}$.

\Paragraph{CNOT circuits.}
A \emph{CNOT circuit} $(\Inputs, \Gates)$ is similar to an XOR circuit, except that
each gate has the form $x_j \gets x_j + x_k$ for distinct $x_j, x_k \in \Inputs$,
i.e., it updates $x_j$ in place.
Expressions are defined by executing the gates in order:
initially $\Eval(x)=x$ for every $x\in\Inputs$, and a gate $x_j\gets x_j+x_k$
updates $\Eval(x_j)$ to $\Eval(x_j)+\Eval(x_k)$.
The circuit implements the matrix $M\in\GL{d}$ whose $i$-th row is the final expression $\Eval(x_i)$,
where $\GL{d}$ is the general linear group over $\Field$
(i.e., the set of invertible $d\times d$ binary matrices).

\Paragraph{Distances.}
The \emph{size} $|C|$ of a (XOR/CNOT) circuit $C$ is the number of its gates.
For $A\in\Field^{r\times d}$, $\XORDistance(A)$ is the minimum size of an XOR circuit
on $d$ inputs implementing $A$.
For $A\in\GL{d}$, $\CNOTDistance(A)$ is the minimum size of a CNOT circuit
on $d$ inputs implementing $A$.
Intuitively, a CNOT circuit is an XOR circuit in which, after a gate
$g\gets a+b$, the operand $a$ is not
accessible to later gates, and the rows of the implemented matrix are the $d$ values
that remain accessible at the end.
We thus have:

\smallskip
\begin{proposition}\label{prop:xor_cnot_distances}
$\forall A\in\GL{d}\colon \CNOTDistance(A)\geq\XORDistance(A)$.
\end{proposition}

\Paragraph{Vertex cover.}
Given a graph $G=(V,E)$, a \emph{vertex cover} of $G$ is a set $S\subseteq V$ that covers all edges, i.e., $\bigcup_{u\in S} \bigcup_{\{u,v\}\in E}\{\{u,v\}\}=E$.
We write $\tau(G)=\min(\{|S|\colon S \text{ is a vertex cover of } G\})$.
We say that $S$ \emph{induces a Hamiltonian path} if $G[S]$, the graph obtained from $G$ by restricting to vertex set $S$, has a Hamiltonian path, i.e., a path that visits each vertex exactly once.

\smallskip
\begin{lemma}\label{lem:vc}
Given a graph $G$ and $k\in\Nats$, deciding whether $\tau(G)\leq k$ is $\NP$-hard, even under the promise that
\begin{enumerate*}[label=(\roman*)]
\item $\tau(G)$ is even, and
\item $G$ has a minimum vertex cover that induces a Hamiltonian path.
\end{enumerate*} 
\end{lemma}

\section{The Hardness of CNOT Optimization}\label{sec:results}

We start by recalling the result of \cite{Boyar2013} on the $\NP$-hardness of optimal XOR synthesis.
We then lift their reduction to CNOT circuits, and use it to prove our main result.

\Paragraph{Edge forms.}
Let $G=(V,E)$ be a graph with $n=|V|$ and $m=|E|$.
The \emph{edge forms} of $G$ are the expressions
$\EdgeForms_G=\{ z+x_u + x_v \}_{\{u,v\}\in E}$
over the inputs $\Inputs=\{z\}\cup\{x_u\}_{u\in V}$.
An XOR circuit $C=(\Inputs, \Gates)$ 
\emph{implements} $\EdgeForms_G$ if every edge form equals $\Eval(v)$ for some variable $v$ of $C$.
In matrix form, ordering the inputs as $z, x_1,\dots,x_n$ and the edges as $e_1,\dots,e_m$,
the edge forms are the rows of the matrix $F_G\in\Field^{m\times(n+1)}$ whose row for
$e=\{u,v\}$ has a $1$ exactly in the columns of $z$, $x_u$ and $x_v$.
Thus, $C$ implements $\EdgeForms_G$ iff it implements $F_G$.
The following lemma is due to~\cite{Boyar2013}; we give a simpler proof in \cref{sec:proofs}.

\smallskip
\begin{lemma}[\cite{Boyar2013}]\label{lem:edge_forms}
For any graph $G$ of $m$ edges, $\XORDistance(F_G) = m + \tau(G)$.
\end{lemma}

\Paragraph{Reversible edge forms.}
Moving from XOR to CNOT, we notice that the matrices $F_G$ are not reversible; they are square only if $m=n+1$, and not full rank in general.
The natural way to recover reversibility is to require that the circuit also outputs each individual input $z$ and $x_u$ unchanged.
Since the output wires have to also be input wires in CNOT circuits (i.e., the matrix has to be square),
we add one extra input $y_{\{u,v\}}$ per edge $\{u,v\}$, and write each edge form onto it.
Formally, the inputs now are
$\Inputs = \{z\}\cup \{ x_u \}_{u\in V}\cup \{ y_{\{u,v\}} \}_{\{u,v\}\in E}$.
The \emph{reversible edge forms} of $G$ are the expressions
$\RevEdgeForms_G=\{y_{\{u,v\}} + z+x_u + x_v \}_{\{u,v\}\in E}$.
A CNOT circuit $(\Inputs, \Gates)$ \emph{implements} $\RevEdgeForms_G$
if its final expressions are
$\Eval(z)=z$, $\Eval(x_u)=x_u$ for every $u\in V$, and
$\Eval(y_{\{u,v\}})=y_{\{u,v\}}+z+x_u+x_v$ for every $\{u,v\}\in E$.
In matrix form, the circuit implements the standard reversible embedding of $F_G$ into the general linear group:
\[
B_G=\begin{pmatrix} I_{n+1} & 0\\ F_G & I_m\end{pmatrix}\in\GL{n+1+m}.
\]
That is, $B_G$ is simply the linear oracle for $F_G$, mapping
$(\mathbf{z},\mathbf{x},\mathbf{y})\mapsto(\mathbf{z},\mathbf{x},\mathbf{y}+F_G(\mathbf{z},\mathbf{x}))$.
Note that linear oracles are involutions, and thus $B_G^2=I$.

What is the overhead of computing reversible edge forms via CNOT circuits?
Compared to edge forms above, every edge $\{u,v\}$ now has an additional term $y_{\{u,v\}}$.
This, intuitively, increases even the XOR distance by $m$ (and thus also the CNOT distance, due to \cref{prop:xor_cnot_distances}), because all $y_{\{u,v\}}$ are distinct.
The more interesting fact is that, for graphs $G$ satisfying the promises of \cref{lem:vc}, the CNOT distance exceeds $\XORDistance(F_G)$ by at most $m+1$.

\smallskip
\begin{lemma}\label{lem:rev_edge_forms}
For any graph $G$ of $m$ edges such that $G$ has a minimum vertex cover that induces a Hamiltonian path,
$\XORDistance(B_G) = 2m+\tau(G)$ and
$\CNOTDistance(B_G)\in \{2m + \tau(G), 2m+\tau(G) +1  \}$.
\end{lemma}

\Paragraph{The hardness of optimal CNOT synthesis.}
\cref{lem:vc,lem:rev_edge_forms} directly imply the sought $\NP$-hardness.

\smallskip
\begin{theorem}\label{thm:cnot}
Given $A\in\GL{d}$ and $K\in\Nats$, deciding whether $\CNOTDistance(A)\leq K$ is $\NP$-complete,
even under the promise that $A$ is a linear oracle.
\end{theorem}
\begin{proof}
Membership in $\NP$ holds since every $A\in\GL{d}$ has a circuit of size $O(d^2/\log d)$~\cite{Patel2008}, which can be verified in polynomial time.
For hardness, we reduce from vertex cover under the promises of \cref{lem:vc}.
Given $G=(V,E)$ with $m=|E|$ and $k\in\Nats$, let $k'=k$ if $k$ is even, and $k'=k-1$ otherwise.
Since $\tau(G)$ is even, we have $\tau(G)\leq k$ iff $\tau(G)\leq k'$.
We ask whether $\CNOTDistance(B_G)\leq K$, for $K=2m+k'+1$.
If $\tau(G)\leq k'$, then $\CNOTDistance(B_G)\leq 2m+\tau(G)+1\leq K$ by \cref{lem:rev_edge_forms}.
Otherwise $\tau(G)\geq k'+2$, as both $\tau(G)$ and $k'$ are even, and thus
$\CNOTDistance(B_G)\geq 2m+\tau(G)>K$ by \cref{lem:rev_edge_forms}.
\end{proof}

\Paragraph{$\NP$-hardness with free ancillas/swaps.}
\cref{thm:cnot} easily extends to two practically motivated settings,
in which circuits have free access to an additional resource:~ancillas and swap gates.
In both cases, its proof goes through unchanged, as
\begin{enumerate*}[label=(\roman*)]
\item the CNOT upper bound of \cref{lem:rev_edge_forms} is achieved without either resource, and
\item the CNOT lower bound of \cref{lem:rev_edge_forms} follows from the same lower bound for XOR circuits.
\end{enumerate*}
It thus suffices to observe that both resources are already free in XOR circuits,
which have unbounded storage (every gate writes a fresh variable) and free wiring;
hence a CNOT circuit using either resource cannot beat the XOR distance.

\SubParagraph{$\NP$-hardness with free ancillas.}
Ancillas are additional wires, either clean (initialized to $0$) or dirty (holding an arbitrary value).
Consider a CNOT circuit $C$ that implements $A$ using some ancillas,
and consider the setting where all dirty ancillas are $0$.
A CNOT gate whose control currently holds $0$ has no effect,
and one whose target currently holds $0$ simply copies the value of its control onto it.
We obtain an XOR circuit implementing $A$, of no larger size, by
(i)~discarding both types of gates, where for the second type we replace all further appearances of the target by the control, and
(ii)~replacing every other CNOT gate $t\gets t+c$ by an XOR gate $p\gets t+c$, and replacing all further appearances of $t$ by $p$.

\SubParagraph{$\NP$-hardness with free swaps.}
A common operation in CNOT (and quantum) circuits is swapping the data of two wires $a$ and $b$,
implemented via CNOT as $a\gets a + b$, $b\gets b + a$, $a\gets a+b$.
Recent works consider CNOT optimization with free swaps~\cite{Webster2026} or free output permutations~\cite{Meijer_van_de_Griend_2023,Shaik2024}:
circuits may contain SWAP gates $a\leftrightarrow b$, but these do not count on the circuit size,
or the circuit output is a permutation of the required one.
Viewing such a circuit as an XOR circuit (as in \cref{prop:xor_cnot_distances}),
a SWAP gate $a\leftrightarrow b$ can be removed by exchanging $a$ and $b$ in all subsequent gates.
Moreover, XOR circuits are oblivious to permutations of their outputs.
Thus, free SWAPs and output permutations are irrelevant for the XOR lower bound of \cref{lem:rev_edge_forms}.

Note that, with free swaps, connectivity constraints are irrelevant, as  a CNOT between non-adjacent wires can be performed by first swapping them until they become adjacent.
Thus $\NP$-hardness holds even when restricting CNOT gates between wires that are adjacent, according to any (connected) topology, e.g., the line.

\section{Proofs}\label{sec:proofs}

\begin{figure*}[t]
\centering
\tikzset{elimfig/.style={x=0.95cm, y=1cm,
    gv/.style={circle, draw, semithick, fill=white, minimum size=4.6mm, inner sep=0pt, font=\scriptsize},
    gvm/.style={gv, fill=orange!35},
    gedge/.style={semithick},
    map/.style={-{Stealth[length=1.6mm]}, semithick},
    hdr/.style={font=\small},
    note/.style={font=\scriptsize, anchor=west, text=black!70},
    gate/.style={rectangle, rounded corners=1.5pt, draw, semithick, minimum size=4.2mm, inner sep=1pt, font=\scriptsize},
    ggate/.style={gate, fill=blue!12},
    fgate/.style={gate, fill=violet!18},
    igate/.style={gate, fill=black!10},
    zgate/.style={gate, fill=white},
    expr/.style={font=\scriptsize, anchor=west, inner sep=1pt},
  }}
\begin{minipage}[b]{0.495\textwidth}
\centering
\begin{tikzpicture}[elimfig]
  \node[hdr] at (0.45, 1.65) {$G^i$};
  \draw[map] (0.8, 1.65) -- node[above, font=\scriptsize] {delete $u$} (1.95, 1.65);
  \node[hdr] at (2.35, 1.65) {$G^{i+1}$};
  \node[hdr] at (5.4, 1.65) {$C^i$};
  \draw[map] (5.8, 1.65) -- node[above, font=\scriptsize] {$x_u\mapsto z$} (7.85, 1.65);
  \node[hdr] at (8.3, 1.65) {$C^{i+1}$};
  \node[gv]  (v1) at (0,    0.95) {$v$};
  \node[gv]  (v2) at (0,    0.05) {$v'$};
  \node[gvm] (u)  at (0.9,  0.5)  {$u$};
  \draw[gedge] (u) -- (v1);
  \draw[gedge] (u) -- (v2);
  \draw[map] (1.25, 0.5) -- (1.95, 0.5);
  \node[gv] at (2.3, 0.95) {$v$};
  \node[gv] at (2.3, 0.05) {$v'$};
  \node[note] at (2.65, 0.95) {$\tau$ drops by $\leq 1$};
  \node[note] at (2.65, 0.05) {$-1$ edge per $v$};
  \node[ggate, label={[expr]right:$x_u{+}z$}] at (5.4, 0.95) {$g$};
  \draw[map] (7.25, 0.95) -- (7.95, 0.95);
  \node[zgate] at (8.3, 0.95) {$0$};
  \node[fgate, label={[expr]right:$z{+}x_u{+}x_v$}] at (5.4, 0.05) {};
  \draw[map] (7.25, 0.05) -- (7.95, 0.05);
  \node[igate] at (8.3, 0.05) {$x_v$};
\end{tikzpicture}

\smallskip
{\small (a) $\Eval(g)=x_u+z$: delete $u$, and project $x_u\mapsto z$.}
\end{minipage}
\hfill
\begin{minipage}[b]{0.495\textwidth}
\centering
\begin{tikzpicture}[elimfig]
  \node[hdr] at (0.45, 1.65) {$G^i$};
  \draw[map] (0.8, 1.65) -- node[above, font=\scriptsize] {merge} node[below, font=\scriptsize] {$u$ into $w$} (1.95, 1.65);
  \node[hdr] at (2.35, 1.65) {$G^{i+1}$};
  \node[hdr] at (5.4, 1.65) {$C^i$};
  \draw[map] (5.8, 1.65) -- node[above, font=\scriptsize] {$x_u\mapsto x_w$} (7.85, 1.65);
  \node[hdr] at (8.3, 1.65) {$C^{i+1}$};
  \node[gv] (u1) at (0, 0.7) {$u$};
  \node[gv] (w1) at (0.9, 0.7) {$w$};
  \draw[gedge] (u1) -- node[above=0.5pt, font=\tiny, inner sep=1pt] {?} (w1);
  \draw[map] (1.25, 0.7) -- (1.95, 0.7);
  \node[gvm] at (2.3, 0.7) {$w$};
  \node[note] at (2.65, 0.95) {$\tau$ drops by $\leq 1$};
  \node[note] at (2.65, 0.4) {edges drop by $\leq 1$};
  \node[gv] (v3) at (0.45, -0.25) {$v$};
  \node[gv] (u3) at (0,   -1.0)  {$u$};
  \node[gv] (w3) at (0.9, -1.0)  {$w$};
  \draw[gedge] (u3) -- (v3);
  \draw[gedge] (w3) -- (v3);
  \draw[map] (1.25, -0.625) -- (1.95, -0.625);
  \node[gv]  (v3b) at (2.3, -0.25) {$v$};
  \node[gvm] (w3b) at (2.3, -1.0)  {$w$};
  \draw[gedge] (w3b) -- (v3b);
  \node[note] at (2.65, -0.625) {$-1$ edge per $w$};
  \node[ggate, label={[expr]right:$x_u{+}x_w$}] at (5.4, 0.95) {$g$};
  \draw[map] (7.25, 0.95) -- (7.95, 0.95);
  \node[zgate] at (8.3, 0.95) {$0$};
  \node[fgate, label={[expr]right:$z{+}x_u{+}x_w$}] at (5.4, 0.4) {};
  \draw[map] (7.25, 0.4) -- (7.95, 0.4);
  \node[igate] at (8.3, 0.4) {$z$};
  \node[fgate, label={[expr]right:$z{+}x_u{+}x_v$}] at (5.4, -0.25) {};
  \node[fgate, label={[expr]right:$z{+}x_w{+}x_v$}] at (5.4, -1.0) {};
  \draw[map] (7.25, -0.625) -- (7.95, -0.625);
  \node[fgate, label={[font=\scriptsize, inner sep=1pt]below:$z{+}x_w{+}x_v$}] at (8.3, -0.625) {};
\end{tikzpicture}

\smallskip
{\small (b) $\Eval(g)=x_u+x_w$: merge $u$ into $w$, and project $x_u\mapsto x_w$.}
\end{minipage}
\caption{The two cases of the transformation $\langle C^i, G^i\rangle\to\langle C^{i+1}, G^{i+1}\rangle$ in the proof of \cref{lem:edge_forms}.
}
\label{fig:elim_cases}
\refstepcounter{subfig}\label{fig:elim_case_z}%
\refstepcounter{subfig}\label{fig:elim_case_w}%
\end{figure*}
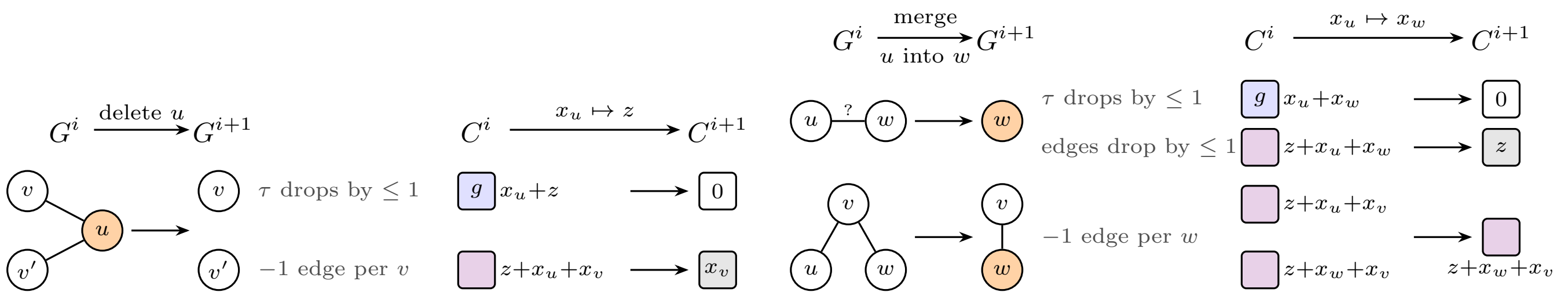

\begin{proof}[\bf Proof of \cref{lem:vc}]
We reduce from vertex cover without the promises, i.e., deciding whether $\tau(H)\leq k$ for a graph $H$ of $n$ vertices, and $k\in\Nats$.
We construct $G$ as two copies of $H$, named $H^1$ and $H^2$, together with a clique $K_{2n}$, where every vertex of $K_{2n}$ further connects to every vertex of $H^1$ and $H^2$.
We argue that $\tau(G)=2\tau(H) + 2n$, and that a minimum vertex cover of $G$ induces a Hamiltonian path.
Then $G$ satisfies both promises, and $\tau(H)\leq k$ iff $\tau(G)\leq 2k+2n$, completing the proof.

Let $U=\{u_1,\dots, u_{2n} \}$ be the vertices of $K_{2n}$.
For $i\in [2]$, let $S^i$ be a minimum vertex cover of $H^i$.
We construct a vertex cover of $G$ as $S=S^1\cup S^2 \cup U$, thus $|S| = 2\tau(H) + 2n$.
Letting $S^1\cup S^2=\{ s_1,\dots, s_t \}$, where $t=2\tau(H)\leq 2(n-1)$, $G[S]$ has the Hamiltonian path
$
s_1, u_1, s_2, u_2,\dots, s_t, u_t, u_{t+1},\dots, u_{2n}
$.

It remains to argue that $S$ is a minimum vertex cover of $G$.
Let $S'$ be a minimum vertex cover of $G$, and we argue that $|S'|\geq 2\tau(H) + 2n$.
Any cover of $K_{2n}$ must contain at least $2n-1$ vertices, and thus $|U\setminus S'|\leq 1$.
If $|U\setminus S'|=1$, then $S'$ must contain all other vertices of $G$, and thus $|S'|=4n-1> 2\tau(H) + 2n=|S|$, since $\tau(H)\leq n-1$, contradicting the minimality of $S'$.
Thus $U\subseteq S'$.
Since $S'$ must also cover the edges of $H^1$ and $H^2$, it must have at least $\tau(H)$ vertices from each, thus $|S'| \geq 2\tau(H) + 2n$.
The desired result follows.
\end{proof}

\Paragraph{The gate elimination method.}
The lower bounds of \cref{lem:edge_forms,lem:rev_edge_forms} are based on \emph{gate elimination}.
Consider an XOR circuit $C$ implementing some matrix $A$,
and an input $x$.
If we \emph{project} $x\mapsto 0$ (resp.\ $x\mapsto y$, for another input $y$), i.e., feed $C$ the value $x=0$ (resp.\ $x=y$), $C$ computes the \emph{restriction}
of every expression to the subspace $x=0$ (resp.\ $x=y$), i.e., every expression with $x$ removed (resp.\ replaced by $y$).
Moreover, every gate $g$ that now evaluates to $0$, to an input, or to the same expression as an earlier gate, has become redundant, and can be removed, replacing its uses by that input or gate.
E.g., if $g$ evaluates to $0$, every later gate $h\gets g+c$ is replaced by $c$ (and thus also removed).
This results in a smaller circuit $C'$ with $|C'| = |C|-r$ gates,
where $r$ is the number of deleted gates.
The process can be repeated on $C'$, projecting further inputs,
until the remaining circuit is simple enough to bound directly.

This argument may look suspicious at first:~it appears to rest on an unproven assumption that the cost of implementing $A$
splits into the cost of a restriction to a subspace plus the cost of projecting to it.
This is not the case:~the projection is merely the vocabulary to describe how to obtain $C'$ from $C$.
In other words, $C'$ could equally be defined purely syntactically, by deleting $x$ and the gates $g$ above;
the projection simply makes it easier to see what $C'$ computes.

\begin{proof}[\bf Proof of \cref{lem:edge_forms}]
We prove that $\XORDistance(F_G)\geq m + \tau(G)$ and $\XORDistance(F_G)\leq m + \tau(G)$.

\SubParagraph{\underline{$\XORDistance(F_G)\geq m + \tau(G)$}.}
Consider any XOR circuit $C=(\Inputs, \Gates)$ implementing $F_G$, and we argue that $|C|\geq m + \tau(G)$.
We define a sequence of transformations
\[
\langle C^0, G^0 \rangle \to \langle C^1, G^1 \rangle \to \dots \to \langle C^q, G^q \rangle
\]
with the following properties.
\begin{invariants}
\item\label{item:inv1} Each $G^i=(V^i, E^i)$ is a graph and each $C^i=(\Inputs^i, \Gates^i)$ is an XOR circuit implementing the edge forms of $G^i$.
\item\label{item:inv2} $C^0=C$, $G^0=G$, and $G^q$ has no edges.
\item\label{item:inv3} For all $0\leq i<q$,
\[
|C^i| - |C^{i+1}| \geq (|E^i| - |E^{i+1}|)  + (\tau(G^i) - \tau(G^{i+1}))\ .
\]
\end{invariants}
Summing \cref{item:inv3} over all $i$, both sides telescope to
\[
|C^0| - |C^{q}| \geq (|E^0| - |E^{q}|) + (\tau(G^0) - \tau(G^q) )\ .
\]
\Cref{item:inv2} implies that $\tau(G^q)=|E^q|=0$, and since $|C^{q}|\geq 0$, we obtain $|C^0| \geq |E^0|  + \tau(G^0)$.
Using \cref{item:inv2} again, we thus obtain $|C|\geq m + \tau(G)$, as desired.

It remains to define the transformation.
By the choice of $C$ and \cref{item:inv2}, \cref{item:inv1} holds for $i=0$.
We now describe the transformation $\langle C^i, G^i \rangle \to \langle C^{i+1}, G^{i+1} \rangle$.
If $G^i$ has no edges, we are done.
Otherwise, since $C^i$ implements the edge forms of $G^i$, there is a gate in $C^i$
whose two operands evaluate to distinct inputs (i.e., some $x_u$, and either $z$ or some $x_v$).
Call the first such gate $g\gets a + b$, and let $\Eval(a)=x_u$, for some $u\in V^i$.
We distinguish the following cases (see \cref{fig:elim_cases}).

\begin{compactitem}
\item $\Eval(b)=z$ (\cref{fig:elim_case_z}).
We obtain $G^{i+1}$ by removing from $G^i$ the vertex $u$ and all its edges.
We obtain $C^{i+1}$ by projecting $x_u\mapsto z$, i.e., $\Inputs^{i+1}=\Inputs^i \setminus \{ x_u \}$, and every use of $x_u$ is replaced by $z$.
Now $\Eval(g)=0$, thus we remove $g$.
Moreover, for every edge $\{u,v\}\in E^i$, $C^i$ has a gate $g_{\{u,v\}}$ evaluating to $z+x_u+x_v$;
these gates are pairwise distinct and distinct from $g$, as they evaluate to distinct expressions in $C^i$.
In $C^{i+1}$, each $g_{\{u,v\}}$ evaluates to the input $x_v$, thus we remove it,
replacing its uses by $x_v$.
Every edge of $G^{i+1}$ is an edge of $G^i$ not incident to $u$, so its form does not contain $x_u$ and is still computed by the same gate;
thus $C^{i+1}$ implements the edge forms of $G^{i+1}$.\\

Note that $\tau(G^i)-\tau(G^{i+1}) \leq 1$, as any vertex cover of $G^{i+1}$ becomes a vertex cover of $G^i$ by adding $u$ to it;
this decrement is covered by the removal of $g$.
Moreover, $|E^i| - |E^{i+1}| = |\{\{u,v\}\in E^i\}|$, which equals the number of removed gates $g_{\{u,v\}}$.
Thus, the inequality of \cref{item:inv3} holds.\\

\item $\Eval(b)=x_w$, for some $w\in V^i\setminus\{u\}$ (\cref{fig:elim_case_w}).
We obtain $G^{i+1}$ by merging $u$ into $w$, i.e., we remove $u$ from $G^i$ and we attach each of its edges to $w$ (if not already present).
We obtain $C^{i+1}$ by projecting $x_u\mapsto x_w$, i.e., $\Inputs^{i+1}=\Inputs^i \setminus \{ x_u \}$, and every use of $x_u$ is replaced by $x_w$.
Now $\Eval(g)=0$, thus we remove $g$.
For every edge $\{u,v\}\in E^i$ with $v\neq w$, the gate $g_{\{u,v\}}$ evaluates to $z+x_w+x_v$ in $C^{i+1}$, and thus implements the form of the (possibly new) edge $\{w,v\}$.
If moreover $\{w,v\}\in E^i$, then $g_{\{u,v\}}$ and $g_{\{w,v\}}$ now evaluate to the same expression,
thus we remove the later one (in order of appearance in $\Gates^i$), replacing its uses by the earlier one.
Finally, if $\{u,w\}\in E^i$, the gate $g_{\{u,w\}}$ now evaluates to the input $z$, thus we remove it,
replacing its uses by $z$.
As before, the removed gates are pairwise distinct, as they evaluate to distinct expressions in $C^i$.
Every other edge of $G^{i+1}$ is an edge of $G^i$ not incident to $u$, so its form is still computed by the same gate;
thus $C^{i+1}$ implements the edge forms of $G^{i+1}$.\\

Note that $\tau(G^i)-\tau(G^{i+1}) \leq 1$, as any vertex cover of $G^{i+1}$ becomes a vertex cover of $G^i$ by adding $u$ to it:
in $G^i$, $u$ covers its own edges, and all other edges of $G^i$ are also edges of $G^{i+1}$.
This decrement is covered by the removal of $g$.
Moreover, the edges of $G^i$ that have no counterpart in $G^{i+1}$ are
$\{u,w\}$ (if present), and one edge for every $v$ adjacent to both $u$ and $w$
(as $\{u,v\}$ and $\{w,v\}$ merge into $\{w,v\}$).
Thus $|E^i| - |E^{i+1}|$ equals the number of removed gates of the form $g_{\{u,w\}}$, $g_{\{u,v\}}$, or $g_{\{w,v\}}$.
Thus, the inequality of \cref{item:inv3} holds.
\end{compactitem}
The above process also extracts a vertex cover of $G$ of size $\tau(G)$, simply by taking $u$ in the cover in each step.

\SubParagraph{\underline{$\XORDistance(F_G)\leq m + \tau(G)$}.}
We construct an XOR circuit $C=(\Inputs, \Gates)$ with $\Inputs = \{z\}\cup \{ x_u \}_{u\in V}$ that implements $F_G$, and has size $|C|=m+\tau(G)$.
Let $S$ be a minimum vertex cover of $G$, thus $|S|=\tau(G)$.
Since $S$ is a vertex cover, we can assign every edge $\{u,v\}\in E$ to an endpoint $\pi(\{u,v\})\in S$.
The circuit proceeds as follows.
\begin{compactenum}
\item\label{step:xor_helpers} For every $v\in S$: $t_v\gets z+x_v$.
\item\label{step:xor_edges} For every  $\{u,v\}\in E$ with $\pi(\{u,v\})=v$: $f_{\{u,v\}}\gets t_v+x_u$.
\end{compactenum}
It follows that $C$ implements $F_G$ and $|C| = m+\tau(G)$.
\end{proof}

\begin{proof}[\bf Proof of \cref{lem:rev_edge_forms}]
The bound $\XORDistance(B_G)\leq 2m+\tau(G)$ follows from a trivial extension to the algorithm in the proof of \cref{lem:edge_forms}, where each gate $f_{\{u,v\}}\gets t_v+x_u$ is followed by one more gate $f'_{\{u,v\}}\gets f_{\{u,v\}}+y_{\{u,v\}}$.
Next, we prove the lower bound $\XORDistance(B_G)\geq 2m + \tau(G)$, which extends to $\CNOTDistance$ due to \cref{prop:xor_cnot_distances}, and the upper bound $\CNOTDistance(B_G)\leq 2m + \tau(G)+1$.

\SubParagraph{\underline{$\XORDistance(B_G)\geq 2m + \tau(G)$}.}
Consider any XOR circuit $C$ implementing $B_G$, and fix an edge $\{u,v\}\in E$.
Since $C$ computes $y_{\{u,v\}}+z+x_u+x_v$, some gate contains $y_{\{u,v\}}$; let $g\gets a+b$ be the first such gate.
Then one of its operands, say $a$, is the input $y_{\{u,v\}}$ itself.
Project $y_{\{u,v\}}\mapsto 0$ and simplify the circuit as in the gate elimination method.
In particular, $g$ now evaluates to $\Eval(b)$, thus we remove it, replacing its uses by $b$.
Repeat this process over all $\{u,v\}\in E$, and let $C'$ be the resulting circuit.
Note that $|C'|\leq |C|-m$, since we have removed at least one gate per $\{u,v\}\in E$.
Moreover, $C'$ implements $F_G$, and by \cref{lem:edge_forms}, we have $|C'|\geq m+\tau(G)$.
We thus arrive at $|C|\geq 2m + \tau(G)$.

\SubParagraph{\underline{$\CNOTDistance(B_G)\leq 2m + \tau(G) + 1$}.}
We construct a CNOT circuit $C$ implementing $B_G$, with $|C|=2m+\tau(G) + 1$.
Let $S$ be a minimum vertex cover of $G$ inducing a Hamiltonian path $s_1,\dots, s_k$,
where $k=\tau(G)$.
We can assign every edge $\{u,v\}\in E$ not on the path
to an endpoint $\pi(\{u,v\})\in S$.
The circuit only targets the wires $z$ and $y_{\{u,v\}}$, and proceeds as follows (see also \cref{fig:helper_chain}).
\begin{compactenum}
\item\label{step:first} $z\gets z+x_{s_1}$.
\item\label{step:loop} For $i=1,\dots,k$:
  \begin{compactenum}
  \item\label{step:other} For every edge $\{s_i,u\}$ not on the path with $\pi(\{s_i,u\})=s_i$:
    $y_{\{s_i,u\}}\gets y_{\{s_i,u\}}+z$ and $y_{\{s_i,u\}}\gets y_{\{s_i,u\}}+x_u$.
  \item\label{step:path} If $i<k$:
    $z\gets z+x_{s_{i+1}}$, \ $y_{\{s_i,s_{i+1}\}}\gets y_{\{s_i,s_{i+1}\}}+z$, and $z\gets z+x_{s_i}$.
  \end{compactenum}
\item\label{step:last} $z\gets z+x_{s_k}$.
\end{compactenum}

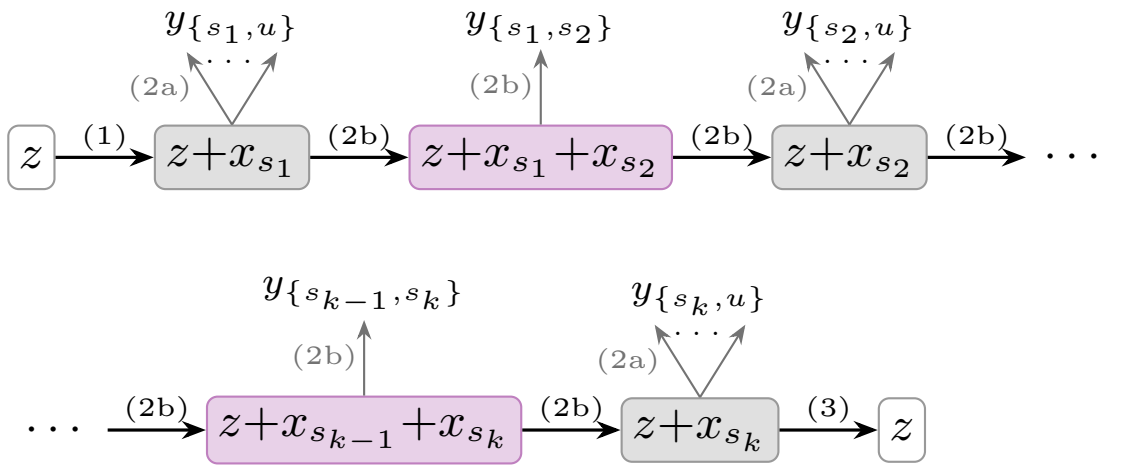
\begin{figure}[htb]
\centering
\begin{tikzpicture}[
    node distance=6.5mm,
    val/.style={rectangle, rounded corners=2pt, draw=black!40, inner sep=2pt,
                minimum height=4.2mm, font=\small},
    zval/.style={val, fill=white},
    hval/.style={val, fill=black!12},
    fval/.style={val, fill=violet!18, draw=violet!50},
    outv/.style={font=\scriptsize, inner sep=1pt},
    zstep/.style={-{Stealth[length=1.6mm]}, semithick},
    ostep/.style={-{Stealth[length=1.3mm]}, black!55},
    lab/.style={font=\tiny, inner sep=1pt},
  ]
  \newcommand{\fan}[1]{%
    \draw[ostep] (#1.north) -- ++(-3mm,4.8mm);
    \draw[ostep] (#1.north) -- ++(3mm,4.8mm);
    \node[font=\tiny, inner sep=0pt] at ($(#1.north)+(0,4.2mm)$) {$\cdots$};
    \node[lab, left, text=black!55] at ($(#1.north)+(-2.2mm,2.4mm)$) {(2a)};
  }
  \node[zval] (a0) {$z$};
  \node[hval, right=of a0] (a1) {$z{+}x_{s_1}$};
  \node[fval, right=of a1] (a2) {$z{+}x_{s_1}{+}x_{s_2}$};
  \node[hval, right=of a2] (a3) {$z{+}x_{s_2}$};
  \node[right=of a3] (a4) {$\cdots$};
  \draw[zstep] (a0) -- node[lab, above] {(1)} (a1);
  \draw[zstep] (a1) -- node[lab, above] {(2b)} (a2);
  \draw[zstep] (a2) -- node[lab, above] {(2b)} (a3);
  \draw[zstep] (a3) -- node[lab, above] {(2b)} (a4);
  \node[outv, above=5mm of a1] (oa1) {$y_{\{s_1,u\}}$};
  \node[outv, above=5mm of a2] (oa2) {$y_{\{s_1,s_2\}}$};
  \node[outv, above=5mm of a3] (oa3) {$y_{\{s_2,u\}}$};
  \fan{a1}
  \draw[ostep] (a2) -- node[lab, left] {(2b)} (oa2);
  \fan{a3}
  \node[below=14mm of a0.south west, anchor=north west] (b0) {$\cdots$};
  \node[fval, right=of b0] (b1) {$z{+}x_{s_{k-1}}{+}x_{s_k}$};
  \node[hval, right=of b1] (b2) {$z{+}x_{s_k}$};
  \node[zval, right=of b2] (b3) {$z$};
  \draw[zstep] (b0) -- node[lab, above] {(2b)} (b1);
  \draw[zstep] (b1) -- node[lab, above] {(2b)} (b2);
  \draw[zstep] (b2) -- node[lab, above] {(3)} (b3);
  \node[outv, above=5mm of b1] (ob1) {$y_{\{s_{k-1},s_k\}}$};
  \node[outv, above=5mm of b2] (ob2) {$y_{\{s_k,u\}}$};
  \draw[ostep] (b1) -- node[lab, left] {(2b)} (ob1);
  \fan{b2}
\end{tikzpicture}
\caption{The values of the $z$ wire, and the outputs they update.}
\label{fig:helper_chain}
\end{figure}

Each reversible edge form $y_{\{u,v\}}+z+x_u+x_v\in\RevEdgeForms_G$
is computed in \cref{step:path} if $\{u,v\}$ lies on the Hamiltonian path, and in \cref{step:other} otherwise.
Moreover, the $x_u$ wires are never targeted, and the $z$ wire is restored to $z$
by \cref{step:last}, so $C$ implements $B_G$.

\Cref{step:first,step:last} use one gate each,
\cref{step:path} uses $3$ gates for each of the $k-1$ path edges,
and \cref{step:other} uses $2$ gates for each of the remaining $m-k+1$ edges.
Thus $|C| = 2 + 3(k-1) + 2(m-k+1) =  2m+\tau(G)+1$.
\end{proof}

\bibliography{refs}
\end{document}